\documentclass{iopart}
\usepackage{iopams}
\usepackage{setstack}
\usepackage{graphicx}
\usepackage{cite}

\newcommand{\commentout}[1]{}

\newcommand{\cre}[1]{c_{#1}^\dagger}
\newcommand{\ann}[1]{c_{#1}^{\phantom\dagger}}
\newcommand{\des}{\ann}

\newcommand{\sgn}{\mathop{\mathrm{sgn}}}

\newcommand{\D}{\;\mathrm{d}}

\newcommand{\bk}{\mathbf{k}}

\newcommand{\I}{\mathrm{i}}

\newcommand{\util}{\tilde U}

\newcommand{\fcre}[1]{f^\dagger_{#1\vphantom)}}
\newcommand{\fdes}[1]{f^{\phantom\dagger}_{#1\vphantom)}}

\renewcommand{\Im}{\mathop{\mathrm{Im}}}
\renewcommand{\Re}{\mathop{\mathrm{Re}}}

\newcommand{\ie}{\emph{i.e.}}
\newcommand{\cf}{\emph{cf.}}
\newcommand{\eg}{\emph{e.g.}}
\newcommand{\vs}{\emph{vs.}}

\newcommand{\viz}{\emph{viz.}}

\newcommand{\boxcomment}[1]{}

\newcommand{\half}{\case{1}{2}}

\newcommand{\omegam}{\omega_\mathrm{m}}

\newcommand{\srefs}[2]{secs.\ (\ref{#1},\ref{#2})}
\newcommand{\erefs}[2]{(\ref{#1},\ref{#2})}

\begin{document}
\title{A local moment approach to the degenerate Anderson impurity model.}
\author{Martin R. Galpin, Anne B. Gilbert and David E. Logan}
\address{Oxford University, Chemistry Department, Physical and Theoretical Chemistry Laboratory, South Parks Road, Oxford, OX1 3QZ, UK}
\begin{abstract}
The local moment approach is extended to the orbitally-degenerate ($SU(2N)$) Anderson impurity model (AIM). Single-particle dynamics are obtained over the full range of energy scales, focussing here on particle-hole symmetry in the strongly correlated regime where the onsite Coulomb interaction leads to many-body Kondo physics with entangled spin and orbital degrees of freedom. The approach captures many-body broadening of the Hubbard satellites, recovers the correct exponential vanishing of the Kondo scale for all $N$, and its universal scaling spectra are found to be in very good agreement with numerical renormalization group (NRG) results. In particular the high-frequency logarithmic decays of the scaling spectra, obtained here in closed form for arbitrary $N$, coincide essentially perfectly with available numerics from the NRG. A particular case of an anisotropic Coulomb interaction, in which the model represents a system of $N$ `capacitively-coupled' $SU(2)$ AIMs, is also discussed. Here the model is generally characterised by two low-energy scales, the crossover between which is seen directly in its dynamics.
\end{abstract}
\maketitle

%--------------------------------------------------------------------

\section{Introduction}
\label{sec:intro}

The Anderson impurity model (AIM) \cite{hewsonbook} plays a central role in our understanding of correlated-electron physics. It has long been used to explain the properties of a diverse range of systems where strong, local Coulomb interactions are paramount \cite{hewsonbook}, most recently in the area of electronic transport through nanostructures such as semiconducting quantum dots, carbon nanotubes and single-molecule devices \cite{glazmanphysicsworld,kouwenhovenqdreview}.

The original AIM~\cite{anderson} describes a system of spin-$\half$ electrons, tunnelling between a metallic conduction band and a single impurity level on which they experience a local Coulomb repulsion `$U$'. This interaction renders the problem highly non-trivial, but the physics of the model has been uncovered over several decades by a range of techniques, see \eg\
\cite{hewsonbook,ayh,nozieresfl,wilson,kww1,tsvelikreview,bickers,keiterkimball1971,kuramoto1983,tvr,gunnschon,readnewns1983,coleman,affleck,krohawolfle1998,let}. Most significantly, the Kondo effect---the quenching of the impurity spin via formation of a complex many-body state between the impurity and conduction band -- is now rather well understood.

In this paper, we consider the $SU(2N)$ generalisation of the AIM~\cite{hewsonbook}, in which the spin-$\half$ electrons of the original model are replaced by particles with $2N$ internal `flavour' degrees of freedom. As well as being important from a theoretical point of view, the $N=2$ case in particular has direct relevance to recent experiments on carbon nanotube quantum dots \cite{jarillo,makarovski,choinano,busser,fbaprl}. The flavour index of the particles here corresponds to a combination of electron spin and orbital indices, and the model exhibits
an $SU(4)$ Kondo effect in which spin and orbital degrees of freedom are entangled \cite{borda,lopez}.

The $SU(2N)$ AIM has appeared in the literature in many guises over the years. In the $U\to\infty$ limit, where the impurity can be at most singly-occupied, the model becomes somewhat simpler to analyze. It is integrable, and hence can be studied using the Bethe \emph{ansatz} \cite{tsvelikreview,ogievetski,kawakamidegenerate,zvyagin}. 
Various approximations falling under the umbrella of `large-$N$ expansions' \cite{bickers,keiterkimball1971,kuramoto1983,tvr,gunnschon,readnewns1983,coleman} have also been developed. These become exact in the limit $N\to\infty$, where their thermodynamics agree with the Bethe \emph{ansatz} results \cite{hewsonbook}. Generalising large-$N$ techniques to finite-$U$ is however complicated \cite{holmschonhammer,pruschkegrewe,qinkeiter,schillerzevin,holmkreeschonhammer,krohawolfle}, 
 rendering this still an active area of research today \cite{grewereview}. Recent progress has also been made in a different direction using a slave-rotor mean-field theory \cite{slaverotor}, which recovers the correct exponential form of the Kondo scale for arbitrary $N$ at the particle-hole symmetric limit.

In addition to analytical approaches, $SU(2N)$ models have also been studied numerically. Thermodynamics and dynamics have been calculated using quantum Monte Carlo (QMC), see \eg\ \cite{boncagubernatis}. While QMC can in principle handle any degeneracy $N$, interaction strength $U$ and temperature $T$ \cite{boncagubernatis}, the method is computationally rather expensive, and cannot reach the exponentially-small low energy scales on which the Kondo effect is fully manifest. The numerical renormalization group (NRG) \cite{wilson,kww1,bkp} has also of course been applied with great success to the $SU(2)$ model (for detailed reviews see \eg\ \cite{hewsonbook,bkp}),
as well as to finite-$U$ $SU(4)$ AIMs, \eg\ \cite{fbaprl,ddprl,mrgccstat,mrgccdyn,ccgate}. NRG
is numerically exact down to the lowest energy scales, although the scaling with $N$ of its CPU and
computer memory usage is such that at present it is realistically limited in practice to the $SU(2)$ and $SU(4)$ cases.

Here we analyze the $SU(2N)$ model using the `local moment approach' (LMA), considering explicitly 
the particle-hole symmetric limit where the impurity is occupied by $N$ particles. The LMA -- developed originally for the $SU(2)$ AIM \cite{let,mattdel_asym,nigelscalspec,nigelfield,nigeltherm,mrgtse,latestpaper} and subsequently extended to encompass both the pseudogap \cite{mattpseud1,nrglmacomppseud,mattpseud,mattgarethpseud} and gapped \cite{mrggaplma,mrggappt} AIMs, as well as lattice-based models within dynamical mean-field theory \cite{vickipam,rajaki,rajaepjb,rajapamtheory,rajapamexp,anneki} -- has been shown to circumvent many of the traditional deficiencies of previous many-body approaches. It can handle all interaction strengths, from the non-interacting limit up to the `strong-coupling' regime where electron correlations dominate; and recovers the correct physics at both high and low energies, the latter being the Kondo effect described above and its ultimate Fermi-liquid description on the lowest energy scales. It is also physically transparent and computationally straightforward; many of its results can in fact be obtained analytically, including non-trivial closed formulae for single-particle dynamics of the model in the strong-coupling regime. 

We explain in the following that generalisation of the LMA to the $SU(2N)$ AIM is quite natural (and
 does not require significant additional computation when applied to arbitrary $N$). In the strong-coupling regime, $N$ essentially appears only as an additional prefactor to the LMA self-energies, and hence many of the known LMA results for the $SU(2)$ model can be extended quite straightforwardly. In particular, we show that the LMA recovers the correct exponential dependence of the Kondo scale on $N$, and derive a closed expression for the `tails' of the universal $SU(2N)$ single-particle scaling spectrum that agrees excellently with NRG results for the $SU(2)$ and $SU(4)$ cases.

The paper is structured as follows. \Sref{sec:model} defines the $SU(2N)$ AIM, shows its connection to a model of $N$ capacitively-coupled $SU(2)$ AIMs, and defines the particle-hole symmetric case on which we focus 
here. In \sref{sec:lma} we introduce the LMA, starting with an examination of the non-interacting and  `atomic' limits, before moving on to the mean-field level of unrestricted Hartree-Fock and then the LMA itself. \Sref{sec:results} contains a summary of our results for the model, both numerical and analytic as mentioned above, together with direct comparison to NRG results for single-particle dynamics in the $SU(4)$ and $SU(2)$ cases. We end the paper by discussing briefly the physics of the model away from $SU(2N)$ symmetry [\sref{sec:ccresults}], and provide a short conclusion in \sref{sec:conc}.

%
%---------------------------------------------------------------------

\section{Model}
\label{sec:model}
In conventional notation the Hamiltonian for the $SU(2N)$ Anderson impurity model is \cite{hewsonbook}
\begin{equation}
\label{eq:su2nham}
\fl\hat H = \sum_{\bk,m}\epsilon_\bk^{\phantom\dagger}\cre{\bk m}\des{\bk m} + \epsilon \sum_m \hat n_m + \frac{U}{2}\sideset{}{'}\sum_{m,m'}\hat n_{m}\hat n_{m'} + V\sum_{\bk,m}\left(\cre{\bk m}\fdes{m} + \mathrm{h.c.}\right),
\end{equation}
where $\hat n_m = \fcre{m}\fdes{m}$, the flavour index $m$ takes on $2N$ distinct values, and the prime in the third sum means $m'=m$ is excluded. The Hamiltonian thus describes a non-interacting host band of `flavourful' particles (first term), plus a localised impurity orbital with onsite Coulomb repulsion $U$ (second and third terms), with tunnel couplings between the two (final term).

\Eref{eq:su2nham} highlights clearly the inherent $SU(2N)$ symmetry of the model. But it will also be instructive to rewrite the Hamiltonian in a way that reveals its connection to a problem of spinful $SU(2)$ electrons. To this end, we regard $m$ as a composite index, $m\equiv(i,\sigma)$, where $i$ takes on $N$ integral `site' indices from $1$ to $N$ and $\sigma=$ $\uparrow$ or $\downarrow$ represents an $SU(2)$ spin index. Sums over $m$ then become joint sums over $i$ and $\sigma$, and \eref{eq:su2nham} can be rewritten as
\begin{equation}
\label{eq:su2nhamisig}
\fl\hat H = \sum_{\bk,i,\sigma}\epsilon_\bk^{\phantom\dagger}\cre{\bk i\sigma}\des{\bk i\sigma} + \epsilon\sum_{i,\sigma}\hat n_{i\sigma} + \frac{U}{2}\sideset{}{'}\sum_{i,j,\sigma,\sigma'}\hat n_{i\sigma}\hat n_{j\sigma'} + V\sum_{\bk,i,\sigma}\left(\cre{\bk i \sigma}\fdes{i\sigma} + \mathrm{h.c.}\right) 
\end{equation} 
with $\hat n_{i\sigma} = \fcre{i\sigma}\fdes{i\sigma}$, and the prime on the sum means that the term with both $i=j$ \emph{and} $\sigma=\sigma'$ is excluded. The interaction term can be simplified by writing
\begin{equation}
\label{eq:sumsimp}
\sideset{}{'}\sum_{i,j,\sigma,\sigma'} \equiv \sum_{i,j}\sideset{}{'}\sum_{\sigma,\sigma'}\delta_{ij} + \sideset{}{'}\sum_{i,j}\sum_{\sigma,\sigma'}\delta_{\sigma'\sigma}
\end{equation}
(with $\delta_{ab}$ the Kronecker delta), such that
\begin{equation}
\label{eq:su2nhamndots}
\fl\hat H = \sum_i\left[\sum_{\bk,\sigma}\epsilon_\bk^{\phantom\dagger}\cre{\bk i\sigma}\des{\bk i\sigma} + \epsilon \hat n_i + U\hat n_{i\uparrow}\hat n_{i\downarrow} + 
V\sum_{\bk,\sigma}\left(\cre{\bk i \sigma}\fdes{i\sigma} + \mathrm{h.c.}\right)\right] +  \frac{U}{2}\sideset{}{'}\sum_{i,j} \hat n_i \hat n_j
\end{equation} 
where $\hat n_i = \sum_\sigma n_{i\sigma}$. In this form, the Hamiltonian describes $N$ equivalent \emph{capacitively-coupled} $SU(2)$ Anderson models, in the limit of identical onsite and intersite Coulomb interactions. The LMA developed here can in fact be straightforwardly generalised to the case when the intersite interaction is less than $U$; we comment briefly on the main results of such in \sref{sec:ccresults}, but until then focus on the fully-symmetric $SU(2N)$ limit corresponding to \eref{eq:su2nham}.

For relative simplicity we also consider the model in the particle-hole symmetric limit, with a standard flat, infinitely-wide conduction band with density of states $\rho_0$~\cite{hewsonbook}. The latter automatically ensures particle-hole symmetry of the host; the impurity part of \eref{eq:su2nham} under a particle-hole transformation $\fcre{m}\to-\fdes{m}$ becomes 
\begin{equation}
\sum_m \left[-\epsilon - (2N-1)U \right] \hat n_m +  \frac{U}{2}\sideset{}{'}\sum_{m,m'} \hat n_m \hat n_{m'}
\end{equation}
and thus the model is fully particle-hole symmetric when 
\begin{equation}
\label{eq:phsym}
\epsilon = -(N - \half)U.
\end{equation}
All $\epsilon$s that appear in the following are implicitly tied to $U$ in this way. For later use, note that the Hamiltonian \eref{eq:su2nham} in this limit reduces (modulo an irrelevant constant) to the capacitive charging form
\begin{equation}
\label{eq:su2nhamsym}
\hat H = \sum_{\bk,m}\epsilon_\bk^{\phantom\dagger}\cre{\bk m}\des{\bk m} + \frac{U}{2}(\hat N - N)^2 + V\sum_{\bk,m}\left(\cre{\bk m}\fdes{m} + \mathrm{h.c.}\right),
\end{equation} 
with $\hat N=\sum_m \hat n_m$ the \emph{total} impurity charge operator. We add here that the subsequent approach is not in fact restricted to the particle-hole symmetric limit \emph{per se}; the LMA has already been used successfully to describe the $SU(2)$ AIM away from particle-hole symmetry \cite{mattdel_asym}, and generalisation of the approach below to the particle-hole asymmetric $SU(2N)$ model can be developed in a similar vein.  

To examine the dynamics of the $SU(2N)$ AIM, the central quantity of interest is the time-ordered, flavour-$m$ impurity Green function $G_m(\omega)$, defined as usual \cite{fwbook} by the Fourier transform of $G_m(t) = -\I\langle \hat{T} \{ \fdes{m}(t) \fcre{m}(0) \}\rangle$ (with $\hat T$ the Wick time-ordering operator), and
with corresponding single-particle spectrum $D_m(\omega) = -\pi^{-1}\sgn(\omega) \Im G_m(\omega)$. It
is this on which we naturally focus in the paper.

%
%---------------------------------------------------------------------
\section{Local moment approach}
\label{sec:lma}

Before we develop the LMA for the $SU(2N)$ AIM, it is instructive to comment briefly on two special cases: the non-interacting limit of $U=0$ (but arbitrary $V$), and the atomic limit with $V=0$ (but arbitrary $U$). 
While only glimpses of the physics of the full model are seen in these rather trivial limits, our main purpose in considering them is that any credible approximate theory for the model should recover both limits as particular cases.

\subsection{Non-interacting limit}
\label{sec:nil}
In the non-interacting limit, $U=0$, the impurity Green function (here denoted by $g_{m}(\omega)$) is readily
obtained in closed form, in direct analogy to the $SU(2)$ case \cite{hewsonbook}, \viz
\begin{equation}
\label{eq:gni}
g_{m}(\omega) = \frac{1}{\omega^+  - \Delta(\omega)}
\end{equation}
with $\omega^+ = \omega + \I 0^+ \sgn(\omega)$. Here $\Delta(\omega) = \sum_{\bk}|V|^2/(\omega^+ - \epsilon_{\bk})$ is the host-impurity hybridisation function, reducing to
$\Delta(\omega) = -\I\sgn(\omega)\Delta_0$ for the conventional wide-band host considered explicitly, where
$\Delta_0 = \pi\rho_0 |V|^2$ (with $\rho_0$ the conduction band density of states, as above).
The spectrum of \eref{eq:gni} is simply
\begin{equation}
\label{eq:dni}
\pi\Delta_0\, d_m(\omega) = \frac{\Delta_0^2}{\omega^2 + \Delta_0^2};
\end{equation}
the Lorentzian form of which reflects the lifetime broadening (of width $\Delta_0$) of the impurity level due to its hybridization with the host band. This introduces a natural energy scale to the problem, and it is convenient to use $\Delta_0$ as the unit of energy in the rest of the paper (unless stated otherwise). To this end, we define the reduced Coulomb interaction $\tilde U = U/\Delta_0$ and the reduced frequency $\tilde\omega = \omega/\Delta_0$.  

\subsection{Atomic limit}
\label{sec:al}
The atomic limit ($V=0$) is trivially soluble for any value of $U$, since the impurity number operator $\hat N$ commutes with $\hat H$.  The basic physics follows directly from the Hamiltonian in \eref{eq:su2nhamsym}. When $U>0$, the ground state is necessarily degenerate. The impurity is occupied by $N$ particles in its ground state, and these can be distributed amongst the $2N$ orbitals in $W_{N} = {2N\choose N}$ distinct  ways. Single-particle excitations connect the degenerate ground state manifold to further $N\pm 1$-particle degenerate manifolds, all lying $U/2$ higher in energy.

In the absence of any special preparation, the particular ground state adopted by the system is equally likely to be any one of the $W_N$ possibilities. The single-particle excitation spectrum for a given flavour $m$ is then the statistical average of the individual flavour-$m$ spectra of all the $W_N$ ground states. In precisely $W_N/2$ of these, there will be a particle in flavour orbital $m$ and hence\footnote{All energies are measured
with respect to the Fermi level.} a `removal' pole in the spectrum at $\omega = -U/2$. In the other $W_N/2$ cases the $m$th flavour orbital will be unoccupied, leading to an `addition' pole at $\omega = +U/2$. As a result, the 
spectrum in the atomic limit consists of two poles of weight $\frac{1}{2}$ lying at $\omega=\pm U/2$, i.e.
\begin{equation}
\label{eq:gal}
G_m(\omega) \overset{V=0} = \frac{1}{2}\left[\frac{1}{\omega^+ + \half U} + \frac{1}{\omega^+ - \half U}\right],
\end{equation}
independently of $N$.

\subsection{Mean field}
\label{sec:uhf}
Next we turn to an analysis at static mean-field (MF) level.  By analogy to unrestricted Hartree-Fock (UHF)
for the $SU(2)$ case \cite{anderson}, we allow for the most general form of local symmetry-breaking,
writing $\hat n_m = \langle \hat{n}_m \rangle + \delta \hat n_m$ for all flavour indices $m$. Substituting this into \eref{eq:su2nham} and neglecting the fluctuation terms quadratic in $\delta \hat n_m \delta \hat n_{m'}$ gives, to within an irrelevant constant,
\begin{equation}
\label{eq:su2nhamhf}
\hat H = \sum_{\bk,m}\epsilon_\bk^{\phantom\dagger}\cre{\bk m}\des{\bk m} + \sum_m \epsilon_m \hat n_m + V\sum_{\bk,m}\left(\cre{\bk m}\fdes{m} + \mathrm{h.c.}\right),
\end{equation} 
with
\begin{equation}
\label{eq:epsm}
\epsilon_m = \epsilon + U\sum_{m'\ne m}\langle \hat{n}_{m'} \rangle.
\end{equation}
Under a particle-hole transformation $\hat{N} \rightarrow 2N-\hat{N}$, and since $\hat{H}$ is invariant under this transformation (see \eref{eq:su2nhamsym}), particle-hole symmetry implies the single constraint 
\begin{equation}
\label{eq:phsymmfconstraint}
\sum_m  \langle \hat{n}_m \rangle = N~,
\end{equation}
so that there are $2N-1$ possible channels in which symmetry can be broken at MF level.
Use of \eref{eq:phsymmfconstraint} in \eref{eq:epsm}, along with \eref{eq:phsym}, yields an expression for $\epsilon_m$ in terms of $\langle \hat{n}_m\rangle$ alone, \viz
\begin{equation}
\label{eq:epsmm}
\epsilon_m = U\left(\half - \langle \hat{n}_{m} \rangle\right).
\end{equation}

Since the Hamiltonian in \eref{eq:su2nhamhf} is separable for each flavour $m$, the MF impurity Green functions 
follow in direct analogy to \eref{eq:gni} as
\begin{equation}
\label{eq:gmmf}
\mathcal{G}_m(\omega) = \frac{1}{\omega^+ - \epsilon_m + i\sgn(\omega)\Delta_0}~.
\end{equation}
The mean number of flavour-$m$ particles on the impurity is then given by integrating the spectrum $\mathcal{D}_m(\omega) = -\pi^{-1}\sgn(\omega)\Im \mathcal{G}_m(\omega)$ up to the Fermi level: 
\begin{equation}
 \langle \hat{n}_{m} \rangle = \int_{-\infty}^0 \mathcal{D}_m(\omega)\D\omega
=\frac{1}{2} - \frac{1}{\pi}\tan^{-1}\left(\frac{\epsilon_m}{\Delta_0}\right)\label{eq:nm}.
\end{equation}
Together with \eref{eq:epsmm}, this implies the following MF self-consistency equation for each $\langle \hat{n}_{m} \rangle$:
\begin{equation}
\label{eq:nmsc}
\langle \hat{n}_{m} \rangle  = \frac{1}{2} - \frac{1}{\pi}\tan^{-1}\left(\frac{U\left(\half - \langle \hat{n}_{m} \rangle\right)}{\Delta_0}\right)
\end{equation}
subject to the constraint of \eref{eq:phsymmfconstraint}.

The simplest solution of \eref{eq:nmsc} and \eref{eq:phsymmfconstraint} arises when $\langle \hat{n}_{m} \rangle = \half$ for all $m$. This is \emph{restricted} Hartree-Fock (RHF), in which the $SU(2N)$ symmetry of the impurity is left entirely unbroken. In this case, by \eref{eq:epsm}, all the $\epsilon_m$s vanish, and hence \eref{eq:gmmf} reduces to the non-interacting Green function $g_m(\omega)$ [\eref{eq:gni}] for all $m$. But while RHF is a (trivial) self-consistent solution of the MF equations, it is not the solution of lowest energy. By breaking the symmetry of \eref{eq:su2nham}, variationally lower solutions can be generated \cite{anderson}.

We have analysed the broken-symmetry MF solutions in detail. As now shown, the self-consistent solutions 
have the $2N$ mean-field energies $\epsilon_m$ partitioned into two groups of $N$ degenerate levels. We label
the level energies of the two groups as  $\epsilon_+$ and $\epsilon_-$, and the mean occupation of each level as $\langle \hat n_{+} \rangle$ and $\langle \hat n_{-}\rangle$, respectively. For convenience in what follows,
we simply refer to the two groups of $N$ levels as the `$+$' and `$-$' groups, respectively.

Upon partitioning the levels between these two groups, the $2N$ equations in
\eref{eq:nmsc} reduce to just two:
\begin{equation}
\label{eq:nmscssb}
\langle \hat n_{\pm} \rangle  = \frac{1}{2} -
\frac{1}{\pi}\tan^{-1}\left(\frac{U\left(\half - \langle \hat n_{\pm}
\rangle\right)}{\Delta_0}\right),
\end{equation}
and the constraint in \eref{eq:phsymmfconstraint} reads
$\sum_{m}\langle \hat n_{m} \rangle \equiv N\langle \hat n_{+} \rangle +
N\langle \hat n_{-} \rangle= N$, \ie 
\begin{equation}
\label{eq:mfssbcons}
\langle \hat n_{+} \rangle + \langle \hat n_{-} \rangle= 1.
\end{equation}
This pair of self-consistency equations, \eref{eq:nmscssb} and
\eref{eq:mfssbcons}, is formally identical to that of the $SU(2)$ AIM at UHF level
\cite{anderson}. In addition to the RHF solution described above ($\langle
\hat n_\pm \rangle = \half$), a symmetry-broken solution is thus found to arise
\cite{anderson} whenever $\tilde U > \pi$: physically the `$+$' and `$-$'
groups are split about $\omega=0$, leading to a net lowering of the energy.
The splitting of the levels is described most easily by introducing the
`moment' $\mu$, writing $\langle \hat n_{\pm}\rangle = \half(1\mp\mu)$
(which automatically satisfies \eref{eq:mfssbcons}). The remaining
self-consistency equation, \eref{eq:nmscssb}, then reads
\begin{equation}
\label{eq:muscssb}
\mu = \frac{2}{\pi}\tan^{-1}\left(\frac{\tilde U\mu}{2}\right),
\end{equation}
from which it follows that symmetry-broken solutions with $\mu\ne 0$ arise in
degenerate pairs; namely $\mu = +|\mu_0|$ and $\mu = -|\mu_0|$, which we denote
respectively as the `$A$'-type and `$B$'-type solutions. From
\eref{eq:epsmm} one sees that an `$A$'-type solution has the energy levels
at $\epsilon_\pm = \pm\half U|\mu_0|$, while the `$B$'-type solution by
contrast has $\epsilon_\pm = \mp\half U|\mu_0|$.

For a given type of solution ($A$ or $B$), the mean-field ground state is
degenerate.  The $2N$ mean-field impurity levels can be partitioned between
two groups of $N$ levels in $W_N$ ($={2N\choose N}$) distinct ways. For example, one solution
corresponds to allocating flavours $m=1\;$--$\;N$ to the `$-$' group and the
remaining flavours $m=(N+1)\;$--$\;2N$ to the `$+$' group; a second solution
arises when the two sets of flavour components are completely interchanged;
and other solutions can obviously be obtained by interchanging only some of
the flavour components  between the two groups. Bearing this in mind, one can if 
desired enumerate all $W_N$ possible degenerate MF states for any given $N$, labelling them 
using the following notation. For each state, the $2N$ possible flavour components are written
in a row, such that the first $N$ belong to the `$-$' group and the remaining $N$ are those of 
the `$+$' group. For example, in the case $N=2$ [$SU(4)$], the state $(1\;2\;3\;4)$
has  $\epsilon_1=\epsilon_-=\epsilon_2$, and $\epsilon_3=\epsilon_+=\epsilon_4$.  All other 
states can be generated by permutations of the flavour components: \eg\ the permutation $(1\;3\;2\;4)$
gives the state in which $\epsilon_1=\epsilon_-=\epsilon_3$, and $\epsilon_2=\epsilon_+=\epsilon_4$. Note that some permutations lead to physically equivalent states---\eg\ $(1\;3\;2\;4)$ and
$(3\;1\;2\;4)$---since the ordering of levels \emph{within} a group is
irrelevant. To avoid this, one considers only permutations in which the two sets of $N$ integers (corresponding to the `$-$' and `$+$' groups) are each in ascending numerical order; \ie\ the state $(1\;3\;2\;4)$ is allowed, whereas $(3\;1\;2\;4)$ is not. In this way, every distinct MF state is counted once, and once only.

The net MF Green function for flavour $m$, denoted here by $G^0_m(\omega)$, then follows by averaging the broken-symmetry, flavour-$m$ Green functions of the $W_N$ degenerate ground states: 
\begin{equation}
\label{eq:gmave}
G^0_m(\omega) = \frac{1}{W_N}\sum_{P} \mathcal{G}^{\alpha,P}_{m}(\omega)
\end{equation}
where $\mathcal{G}^{\alpha,P}_{m}(\omega)$ is the flavour-$m$ Green function
for MF solution $\alpha$ ($=$ either $A$ or $B$) and a particular
permutation $P$ of the flavour components between the $\epsilon_+$
and $\epsilon_-$ levels. The sum over $P$ is taken over the set of permutations
that enumerates all $W_N$ physically distinct states, as described above. 
\Eref{eq:gmave} can then be simplified by the following
arguments. First, as seen explicitly in \eref{eq:gmmf}, the MF
$\mathcal{G}_m(\omega)$ depends on $\epsilon_m$ only. Therefore, on defining
\begin{equation}
\label{eq:gpm}
\mathcal{G}^\alpha_{\pm}(\omega) = \frac{1}{\omega^+ -\epsilon_\pm
+i\sgn(\omega)\Delta_0},
\end{equation}
we obtain $\mathcal{G}^{\alpha,P}_{m}(\omega) =
\mathcal{G}^\alpha_{\pm}(\omega)$ if the permutation $P$ allocates flavour
component $m$ to the `$\pm$' group, respectively. And by symmetry, when the
$W_N$ degenerate states are enumerated the flavour $m$ will appear $W_N/2$
times in each of the `$+$' and `$-$' groups ($W_N$ is necessarily even
for $N\ge 1$). Combining these results, one obtains
\begin{equation}
\label{eq:gsym1}
{G}^0_{m}(\omega) = \frac{1}{2}\left[\mathcal{G}^\alpha_+(\omega) +
\mathcal{G}^\alpha_-(\omega)\right]
\end{equation}
for either $\alpha=A$ or $B$ (and obviously being independent of $m$, reflecting SU(4) rotational invariance).

Hence, the net MF Green function is equivalently the average of the MF Green functions for the $+$ and $-$
groups, \emph{regardless of how the flavour components are actually
allocated to those groups} to form a particular MF solution. Moreover, since
intercoverting $A$ and $B$ exchanges the energies of the two groups such
that $\mathcal{G}^A_{\pm}(\omega) = \mathcal{G}^B_\mp(\omega)$,
\eref{eq:gsym1} becomes
\begin{equation}
\label{eq:gsym2}
{G}^0_{m}(\omega) = \frac{1}{2}\left[\mathcal{G}^A_\pm(\omega) +
\mathcal{G}^B_\pm(\omega)\right];
\end{equation}
the statistically averaged MF Green function is thus equivalently the
average of the `$A$'- and `$B$'-type Green functions for \emph{either given}
group ($+$ or $-$), again regardless of how the flavour components are partitioned
between the groups. In either case, on taking the imaginary part and using
\eref{eq:gpm}, the average UHF spectrum follows as
\begin{equation}
\label{eq:mfspectrum}
\pi\Delta_0 {D}^0_{m}(\omega) = \frac{1}{2}\left[\frac{\Delta_0^2}{(\omega -
\half U |\mu_0|)^2 + \Delta_0^2} + \frac{\Delta_0^2}{(\omega + \half U
|\mu_0|)^2 + \Delta_0^2}\right],
\end{equation}
consisting therefore of two superimposed Lorentzians (`Hubbard satellites')
of half-width, half-maximum (HWHM) $\Delta_0$. 

Note that when $\Delta_0$ is switched off, the MF spectrum 
\eref{eq:mfspectrum} reduces correctly to that of the atomic limit, itself
obtained from the imaginary part of \eref{eq:gal}. This is seen directly by
taking the limit $\Delta_0\to 0$ in \eref{eq:mfspectrum}: the
self-consistency condition of \eref{eq:muscssb} becomes $\mu =\sgn(\mu)$
[and hence $|\mu_0|=1$], while the Lorentzians become delta-functions at
$\omega=\pm\half U |\mu_0| \equiv \pm\half U$. And when $U=0$, the
non-interacting spectrum, \eref{eq:dni}, is trivially recovered by
\eref{eq:mfspectrum}.  

The UHF approach thus captures both the atomic and non-interacting limits (\srefs{sec:nil}{sec:al}),
and in some sense bridges the two when $U$ and $\Delta_0$ are both finite. But it does not of course capture the full physics of the model. \Eref{eq:mfspectrum} is independent of $N$; in reality, the widths of the Hubbard satellites increase with $N$, since this opens up more relaxation channels which lead to enhanced lifetime broadening \cite{let,bickers} (see also \sref{sec:specoverview} below). Most importantly, the low-energy description provided by \eref{eq:mfspectrum} is entirely wrong. The spectrum of the AIM is well known to possess an exponentially-narrow Kondo resonance straddling the Fermi level~\cite{hewsonbook}. This many-body resonance is simply absent at the static MF level of description, and can only be captured by developing a more sophisticated approach.  

In the next section we describe how the local moment approach can be extended to the $SU(2N)$ AIM. As explained
previously in a number of works (see \eg\ \cite{let,mattdel_asym}), the technique overcomes the intrinsic limitations of static MF theory via inclusion, within an inherent two-self-energy framework, of dynamical self-energy contributions which in physical terms embody tunneling between degenerate MF states; and which, in acting thereby to restore the symmetry broken at pure MF level, lead correctly to recovery of Fermi liquid behaviour on low-energy scales.

  To set up the LMA in practice, it is convenient to work with the `capacitively-coupled'
quantum dot picture of \eref{eq:su2nhamndots}, replacing all $m$ indices by ($i,\sigma$).
And just as the MF Green function $G^{0}_{i\sigma}(\omega)$ (\eref{eq:gsym1} or \eref{eq:gsym2})
is independent of the MF solution with which one chooses to work, the same can readily be shown
to hold within the LMA. We thus choose to work with the particular MF solution in which all
$\sigma =\uparrow$ levels are allocated to the `$-$' group, and all $\sigma =\downarrow$ levels
are allocated to the `$+$' group. Physically, this solution corresponds to uniform charges on
all sites $i$, but broken spin-symmetry; more precisely,  when $|\mu|>0$ the `$A$'-type solution has an excess of
$\uparrow$-spin electrons on every (equivalent) site, while the `$B$'-type solution has an excess of
$\downarrow$-spins. Other partitionings of the flavour components ($i,\sigma$) would of course correspond
to different physical pictures -- such as MF solutions that are spin-symmetric but of broken
\emph{charge} symmetry -- but we reiterate that which is used in practice is irrelevant when calculating the
full rotationally invariant Green function. Writing \eref{eq:gsym1} and \eref{eq:gsym2} in the chosen
charge-symmetric, broken spin-symmetry picture, gives
\begin{eqnarray}
\label{eq:mfg1}
G_{i\sigma}^{0}(\omega)~&=\frac{1}{2}\left[{\cal{G}}_{i\uparrow}^{\alpha}(\omega)
+{\cal{G}}_{i\downarrow}^{\alpha}(\omega)\right] \\
&=\frac{1}{2}\left[{\cal{G}}_{i\sigma}^{A}(\omega)
+{\cal{G}}_{i\sigma}^{B}(\omega)\right]
\label{eq:mfg2}
\end{eqnarray}
(where $\alpha =$ either $A$ or $B$), which will be employed in due course; with symmetries
${\cal{G}}^{A}_{i\sigma}(\omega)={\cal{G}}^{B}_{i-\sigma}(\omega)$, such that from \eref{eq:mfg2}
$G_{i\sigma}^{0}(\omega)$ is independent of spin $\sigma$ (and as such spin-rotationally invariant~\cite{let,latestpaper}),
as well as being independent of $i$ (reflecting the equivalence of the $N$ levels in the capacitively coupled
quantum dot picture).

\subsection{Local moment approach}
\label{sec:lmass}
The MF solutions above provide a natural starting point for perturbation theory within a two-self-energy description \cite{latestpaper,let,mattdel_asym}. We take the unperturbed Hamiltonian to have the same symmetries as that of MF, \ie\ to be of form \begin{equation}
\label{eq:h0alpha}
\hat H_0^\alpha =
\sum_{\bk,i,\sigma}\epsilon_\bk\cre{\bk i\sigma}\des{\bk i\sigma} - \sum_{i,\sigma} \alpha\sigma x \hat n_{i\sigma} + V\sum_{\bk,i,\sigma}\left(\cre{\bk i\sigma}\fdes{i\sigma} + \mathrm{h.c.}\right)
\end{equation} 
for each of $\alpha = A$ or $B$, taking $A\equiv +$ and $B\equiv -$ for notational convenience when using the label $\alpha$; and where $x=U|\mu|/2$, with local moment $|\mu|$ (determined at post-MF level via the symmetry restoration condition intrinsic to the LMA~\cite{mattdel_asym}, and specified at the end of the section).
The perturbation term corresponding to \eref{eq:h0alpha} is $\hat H_1^\alpha = \hat H - \hat H_0^\alpha$, \viz
\begin{equation}
\label{eq:h1alpha}
\hat H_1^\alpha = \sum_{i}U\hat n_{i\uparrow}\hat n_{i\downarrow} + \frac{U}{2}\sideset{}{'}\sum_{i,j}\hat n_i \hat n_j + \sum_{i,\sigma} (\epsilon + \alpha\sigma x) \hat n_{i\sigma},
\end{equation}
and in precise parallel to the MF results of the previous section, the unperturbed Green functions for a $\sigma$-spin electron on orbital $i$ are 
\begin{equation}
\label{eq:scrgdef}
\mathcal{G}^\alpha_{i\sigma}(\omega) = \frac{1}{\omega + \alpha\sigma x + \I\sgn(\omega)\Delta_0}~.
\end{equation}
The corresponding perturbed Green functions can be obtained by diagrammatic perturbation theory in $\hat H_1^\alpha$ \cite{latestpaper}. One naturally obtains the Dyson equation 
\begin{equation}
\label{eq:lmadyson}
G^\alpha_{i\sigma}(\omega) = \frac{1}{\left[\mathcal{G}^\alpha_{i\sigma}(\omega)\right]^{-1} - (\epsilon+\alpha\sigma x) - \tilde\Sigma^\alpha_{i\sigma}(\omega)}
\equiv  \frac{1}{\left[g_{i\sigma}(\omega)\right]^{-1} - \tilde\Sigma^\alpha_{i\sigma}(\omega)}
\end{equation}
which defines the two self-energies $\tilde\Sigma^\alpha_{i\sigma}(\omega)$ (for $\sigma =\uparrow ,\downarrow$) arising for a given $\alpha$ [our definition of $\tilde\Sigma^\alpha_{i\sigma}(\omega)$ excluding the static first-order diagram $(\epsilon + \alpha\sigma x)$ coming from the third term in \eref{eq:h1alpha}]. The full, rotationally invariant impurity Green function then follows using directly analogous arguments to those of the previous section, such that (\emph{cf} \eref{eq:mfg1})
\begin{equation}
\label{eq:lmag}
G_{m}(\omega) = \half \sum_{\sigma^{\prime}} G^\alpha_{i\sigma^{\prime}}(\omega)
\end{equation}
(again independently of $m =(i,\sigma)$). The spin symmetry of the Hamiltonian \eref{eq:h0alpha} implies that
\begin{equation}
\label{eq:gabsym}
G^A_{i\sigma}(\omega)=G^B_{i-\sigma}(\omega)~;
\end{equation}
while particle-hole symmetry implies
\begin{equation}
\label{eq:phsymg}
G_{i\sigma}(\omega)=-G_{i\sigma}(-\omega),
\end{equation}
thus simplifying the calculation of \eref{eq:lmag} in practice.

\begin{figure}
\begin{center}\includegraphics{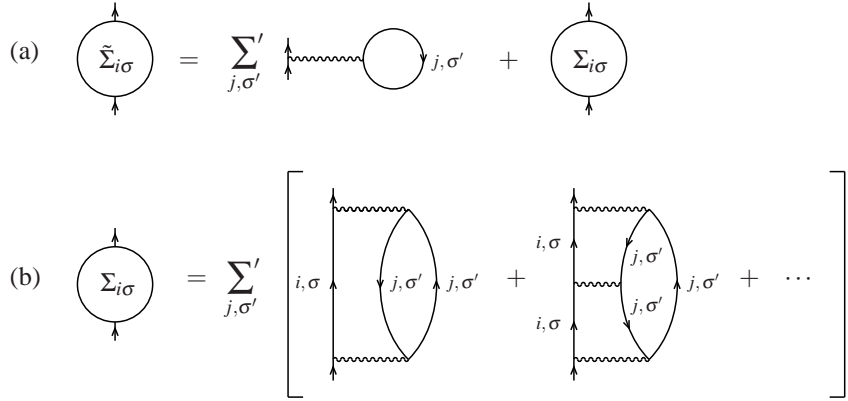}\end{center}
\caption{\label{fig:sesum} (a) Separation of the LMA self-energy $\tilde\Sigma_{i\sigma}(\omega)$ into static plus dynamical terms. The arrowed lines represent the unperturbed Green function $\mathcal{G}_{i\sigma}(\omega)$, and the wavy line represents the Coulomb interaction $U$. (b) Dynamical self-energy approximation within the LMA. In both figures the prime on the summation indicates that the term with $j=i$ \emph{and} $\sigma'=\sigma$ is omitted.}
\end{figure}

\begin{figure}
\begin{center}\includegraphics{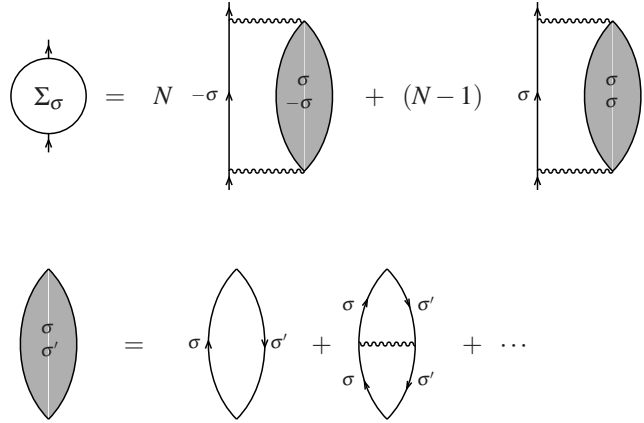}\end{center}
\caption{\label{fig:sigpi} Transformation of \fref{fig:sesum}(b) using the fact that $\mathcal{G}_{j\sigma'}$ is independent of $j$. The shaded bubble represents the RPA spin polarization propagator, \eref{eq:rpapi}.}
\end{figure}

Following \cite{latestpaper,let,mattdel_asym}, we now expand the $\tilde\Sigma^\alpha_{i\sigma}$s perturbatively in their respective $H^\alpha_1$s. For specificity we consider $\alpha=A$ only (and for clarity drop the $A$ superscripts from here on unless indicated otherwise explicitly); the $B$-type self-energy follows (if desired) from the symmetry \eref{eq:gabsym}. As in previous LMA papers (\eg\ \cite{mattdel_asym,let}), we choose to separate the self-energies into static ($\omega$-independent) plus dynamical terms. The former is approximated by the sum over `tadpole' diagrams in \fref{fig:sesum}(a),
 which gives 
\begin{eqnarray}
\label{eq:sigtildestat}
\tilde\Sigma_{i\sigma}(\omega) &= \sideset{}{'}\sum_{j,\sigma'}\half U (1 + \sigma'|\bar\mu|) + \Sigma_{i\sigma}(\omega)\\
&= (N-\half)U - \half \sigma U |\bar\mu| + \Sigma_{i\sigma}(\omega)
\label{eq:sigtildestat2}
\end{eqnarray}
with
\begin{equation}
|\bar\mu| = \int_{-\infty}^0 \D\omega\;\left[\mathcal{D}_{i\uparrow}(\omega) - \mathcal{D}_{i\downarrow}(\omega)\right]
= \frac{2}{\pi}\tan^{-1}\tilde x\label{eq:mubarx}
\end{equation}
(and $\tilde{x}=x/\Delta_{0}$).
%----------------------------------------------
Generalising the approach of \cite{let}, we approximate the dynamical part of the self-energy for a flavour-$m$ [$\equiv (i,\sigma)$] particle by the infinite series shown in \fref{fig:sesum}(b). As we are working from the 
broken-symmetry states in which the  $\mathcal{G}_{j\sigma'}$s are independent of orbital index $j$, all $N$ diagrams describing interactions with $-\sigma$-spin electrons are equivalent, as are all remaining $N-1$ diagrams describing interactions with $\sigma$-spin electrons. The sum over $j,\sigma'$ in \fref{fig:sesum}(b) can thus be replaced by just two terms. Upon recasting the diagrams in terms of random-phase approximation (RPA) polarization propagators, we obtain the result shown in \fref{fig:sigpi}.

Note that all explicit $N$ dependence to the self-energies is contained in the coefficients of the two diagrams,
the broken-symmetry Green functions from which the diagrams are constructed being independent of explicit factors of $N$. In the $SU(2)$ case ($N=1$), only the first diagram on the right-hand side of \fref{fig:sigpi}(a) survives and the self-energy approximation reduces to that used in \cite{let}. When $N>1$, both diagrams contribute,
although we find that the dominant contribution in the strongly correlated regime, $\tilde U\gg 1$, comes from the first diagram alone [see \ref{app:kscale}].

Translating the diagrams in \fref{fig:sigpi} using the Feynman rules gives (\cf\ \cite{let})
\begin{equation}
\label{eq:sigsum}
\Sigma_{i\sigma}(\omega) = N \Sigma^{\sigma-\sigma}_{i\sigma}(\omega) + (N-1) \Sigma^{\sigma\sigma}_{i\sigma}(\omega)
\end{equation}
where
\begin{equation}
\label{eq:sigsigsigdef}
\Sigma^{\sigma\sigma'}_{i\sigma}(\omega) = \frac{U^2}{2\pi\I}\int_{-\infty}^\infty\D\omega_1\; \Pi^{\sigma\sigma'}(\omega_1)\mathcal{G}_{i\sigma'}(\omega_1+\omega).
\end{equation}
The RPA polarization propagators $\Pi^{\sigma\sigma'}(\omega)$ appearing above are translated as 
\begin{equation}
\label{eq:rpapi}
\Pi^{\sigma\sigma'}(\omega) = \frac{{}^0\Pi^{\sigma\sigma'}(\omega)}{1-U\;{}^0\Pi^{\sigma\sigma'}(\omega)},
\end{equation}
with the `bare' polarization bubble
\begin{equation}
\label{eq:pinoughtdef}
{}^0\Pi^{\sigma\sigma'}(\omega) = -\frac{1}{2\pi\I}\int_{-\infty}^\infty\D\omega_1\; \mathcal{G}_{i\sigma'}(\omega_1)\mathcal{G}_{i\sigma}(\omega_1-\omega).
\end{equation}
For the flat-band AIM of interest, the latter can be obtained in closed form. The equations for $^0\Pi^{\sigma-\sigma}(\omega)$ are provided in the Appendix of \cite{let}, and those for $^0\Pi^{\sigma\sigma}(\omega)$ are given in \ref{app:piform} of the present work. 

We note that the numerical calculation of the LMA dynamical self-energies is quite straightforward. In practice it is convenient to determine separately the real and imaginary parts of the various constituents listed above, the former obtained by Hilbert transformation of the latter. As the methods used are essentially identical to those of previous work \cite{let}, we do not discuss them further here.

It remains finally to specify the condition for determination of the local moment $|\mu|$ (entering 
parametrically via $x=U|\mu|/2$ in the Hamiltonian \eref{eq:h0alpha}).
This is achieved by `symmetry-restoration', the central element of the LMA, corresponding physically to restoration
of the symmetry which is broken at pure MF level. It amounts to a single self-consistency condition
on the two self-energies precisely at the Fermi level, $\omega =0$; which, regardless of the specific diagrammatic
approximation used for the self-energies, ensures that the rotationally invariant LMA Green function 
$G_{i\sigma}(\omega)$ recovers Fermi-liquid behaviour on the lowest energy scales \cite{mattdel_asym,latestpaper}. 
Since symmetry restoration is discussed extensively elsewhere
(see \eg\ \cite{mattdel_asym,latestpaper,nigelscalspec,mattpseud,rajaepjb}) we do not elaborate further on it here; but simply note that in direct correspondence with earlier work, the symmetry restoration condition 
here is
\begin{equation}
\label{eq:sr}
\tilde\Sigma_{i\uparrow}(\omega = 0) = \tilde\Sigma_{i\downarrow}(\omega = 0).
\end{equation}
In addition, analyticity of the RPA transverse spin polarization propagator requires \cite{let}
\begin{equation}
\label{eq:stability}
|\mu_0|\le |\mu|=\frac{2x}{U} < 1. 
\end{equation}
Equations~(\ref{eq:sr}) and (\ref{eq:stability}) are readily solved numerically for any choice of the bare model parameters, and are sufficient to determine $x$ uniquely in all cases.

For later use, we also note that a combination of particle-hole symmetry [\eref{eq:phsymg}] and $\uparrow$/$\downarrow$-spin symmetry [\eref{eq:gabsym}] implies that \eref{eq:sr} reduces to
\begin{equation}
\label{eq:srsym}
\epsilon + \tilde\Sigma_{i\sigma}(\omega = 0) = 0. 
\end{equation}
Hence, from \eref{eq:sigtildestat2} and \eref{eq:phsym}, the symmetry restoration condition of \eref{eq:sr} can be written alternatively as
\begin{equation}
\label{eq:srsym2}
\Sigma_{i\sigma}(\omega = 0) = \frac{\sigma}{2} U |\bar\mu|.
\end{equation}

%---------------------------------------------------------------------
\section{Results}
\label{sec:results}
We have analysed in detail the $SU(2N)$ AIM within the LMA described above. Numerical evaluation of the impurity Green function is computationally very inexpensive, and we have examined its behaviour over a wide range of the bare model parameters. Here we describe our main results, focussing in particular on the strong-coupling limit,
$\tilde U =U/\Delta_{0} \gg 1$, where the model enters the Kondo regime in which electron correlations are key.

\subsection{Single-particle spectrum: overview}
\label{sec:specoverview}

\begin{figure}
\begin{center}\includegraphics{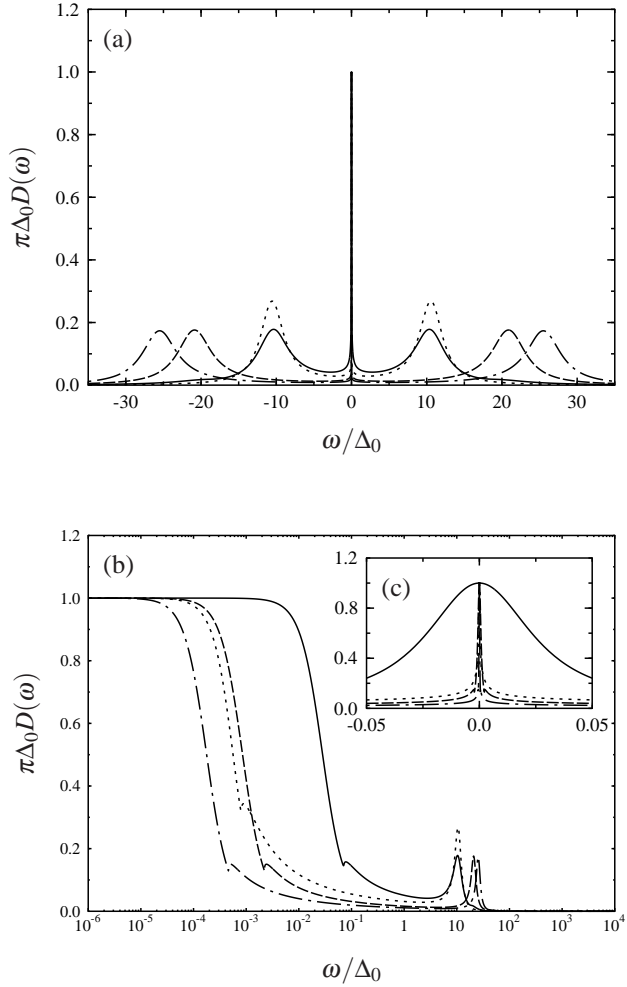}\end{center}
\caption{\label{fig:spec} Representative single-particle spectra of the $SU(2N)$ AIM: $SU(4)$ results are shown for $\tilde U = 20$, $40$ and $50$ (solid, dashed and dot-dashed lines, respectively), and the $SU(2)$ spectrum for $\tilde U=20$ is also shown as a dotted line. (a) shows the behaviour on high frequency scales $\omega\sim U$, while (b) and (c) highlight the low-frequency Kondo resonance.}
\end{figure}

\Fref{fig:spec}(a) shows representative single-particle spectra of the $SU(4)$ AIM in strong-coupling, calculated within the LMA as described in the previous section, for $\tilde U = 20$, $40$ and $50$.
The spectra are plotted as $\pi\Delta_{0}D(\omega)$ \emph{vs} $\omega/\Delta_{0}$ on a linear frequency scale, and 
at the Fermi level in particular are seen to satisfy correctly the dictates of the Freidel sum rule for
all $SU(2N)$ at half-filling, $\pi\Delta_{0}D(\omega =0) =1$~\cite{hewsonbook,fbaprl}.
The figure also shows for comparison the corresponding spectrum of the $SU(2)$ AIM \cite{let} for $\tilde U=20$.

Consider first the high-frequency behaviour seen in \fref{fig:spec}(a). The spectra all show clear Hubbard satellites arising on the scale $|\omega|\simeq U/2$, reflecting physically the single-particle excitations of the atomic limit (see \sref{sec:al}). The finite coupling to the host broadens the Hubbard satellites from the atomic-limit delta-functions in \eref{eq:gal}. This broadening is $\mathcal{O}(\Delta_0)$ in both cases, but the Hubbard satellites of the $SU(4)$ spectra are significantly broader than those of the $SU(2)$ spectrum (moreover, both are broader than the UHF spectrum in \eref{eq:mfspectrum}). The origin of this effect, as alluded to at the end of \sref{sec:uhf},  is many-body scattering from the impurity site \cite{let}, which is
not of course captured at pure MF level. The energy cost to excite particles between different flavour states on the impurity site (\ie\ `flip a spin' in the $SU(2)$ case) is of order the Kondo scale. This is negligible on the energy scale of the Hubbard satellites, and hence the number of relaxation channels contributing to the lifetime broadening of the Hubbard satellites increases with $N$. Indeed in the particle-hole symmetric limit considered here, it is straightforward to show from the LMA that the widths of the Hubbard satellites go as $(N+1)\Delta_0$ in the strong coupling limit, $\tilde U \gg 1$ \cite{bickers,boncagubernatis}.

Moving to lower energy scales, we see that the LMA indeed captures the $SU(4)$ Kondo physics, all spectra in \fref{fig:spec}(a) displaying a sharp Kondo resonance 
centred on the Fermi level $\omega=0$. This behaviour is clearer on the much reduced linear frequency
scale shown in \fref{fig:spec}(c); but is seen most clearly by showing the spectra on a
logarithmic frequency scale as shown in \fref{fig:spec}(b).\footnote{The small spectral `glitch' (seen e.g. at $\tilde\omega \simeq 10^{-1}$ in the solid line of \fref{fig:spec}) is a well known artifact of the RPA ladder sum used within the LMA. While its effect on the spectrum is slight, it can be removed if so desired by the procedure described in \cite{nigelscalspec}.} 
The evident conclusion to be drawn from \fref{fig:spec}(b) is that the low-energy behaviour of the three $SU(4)$ spectra (solid lines) is universal \cite{boncagubernatis,mrgccdyn}. Defining a characteristic Kondo scale $\omega_\mathrm{K}$ by the HWHM of the Kondo resonance, it is quite clear that the three solid lines 
will scale onto each other when plotted on the reduced frequency scale $\omega/\omega_\mathrm{K}$ (provided naturally that $\omega$ itself does not approach non-universal scales of order $|\omega| \sim {\cal{O}}(\Delta_{0})$); as considered in detail in \sref{sec:specscale} below.
 Note further that the $SU(2)$ spectrum shown as a dotted line in \fref{fig:spec}(b), itself known \cite{let,nigelscalspec} to display universal scaling as a function of $\omega/\omega_\mathrm{K}$, does not scale onto the universal $SU(4)$ behaviour: \ie\ the $SU(2)$ and $SU(4)$ scaling spectra are distinct \cite{mrgccdyn} (\cf\ \cite{boncagubernatis}). To understand these results further, we now take a closer look at the low-energy physics of the model.

\subsection{Low-energy scale}
\label{sec:kscale}

Spectral scaling reflects of course the single low-energy scale inherent to the problem in
strong coupling $\util \gg 1$, \viz\ the characteristic $SU(2N)$ Kondo scale. This 
arises naturally within the LMA, in the imaginary part of the polarization propagator $\Pi^{+-}(\omega)$ (see \eref{eq:rpapi}), where it is manifest as a sharp resonance with maximum at $\omega=\omegam \propto \omega_{\mathrm{K}}$ \cite{let}. 

That the LMA $\omegam$ is exponentially small in $\tilde U$ is a direct consequence of the well-known incipient divergence of the RPA ladder sum in \eref{eq:rpapi}, together with self-consistent enforcement of symmetry restoration, \eref{eq:sr}. One can in fact extract the leading $\tilde U \gg 1$ behaviour of the scale analytically,
following \cite{let}. Details of the calculation are given in \ref{app:kscale}, where we obtain the key result
\begin{equation}
\label{eq:kscale}
\omegam \overset{\tilde U \to \infty}\sim c(N) U \exp\left(-\frac{\pi U}{8N\Delta_0}\right),
\end{equation}
with $c(N)$ a constant prefactor for a given $N$. The exponent agrees with numerics \cite{boncagubernatis,ddprl} and a slave-rotor mean-field theory \cite{slaverotor}, and it contains the same (exact) $1/N$ dependence known from analytical studies of the $SU(2N)$ model in the $U\to\infty$ limit \cite{ogievetski,bickers,hewsonbook}. This $1/N$ factor in the exponent means of course that the Kondo scales for $SU(2N)$ models with the same 
$\tilde U\gg 1$ but different $N$ are vastly different, as indeed seen explicitly in \fref{fig:spec}.

\begin{figure}
\begin{center}\includegraphics{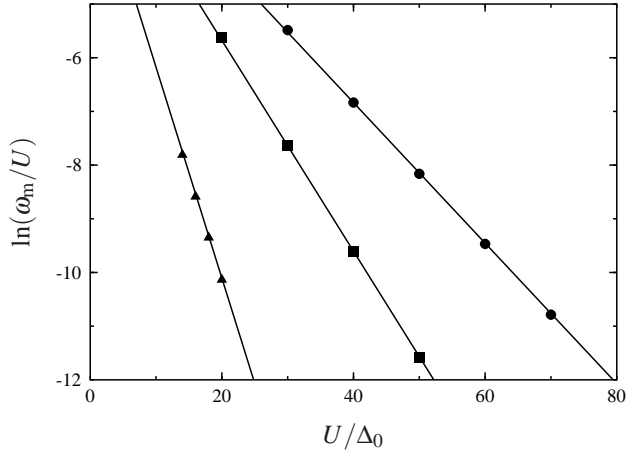}\end{center}
\caption{\label{fig:kscale} The $SU(2N)$ Kondo scale, plotted as $\ln(\omegam/U)$ \vs\ $\tilde U$. The points are numerical results for the $SU(2)$, $SU(4)$ and $SU(6)$ models (triangles, squares and circles, respectively), while the lines are fits to \eref{eq:kscale} for each group.}
\end{figure}

To verify \eref{eq:kscale} itself, we have calculated $\omega_\mathrm{m}$ numerically over a wide range of $\tilde U\gg 1$, for the $SU(2)$, $SU(4)$ and $SU(6)$ cases. The results are shown in \fref{fig:kscale}, plotted as $\ln(\omega_\mathrm{m}/U)$ \vs\ $\tilde U$.
The solid lines are the predictions of \eref{eq:kscale}, with the constant $c(N)$ fit to the numerically determined points (and found in practice to depend rather slowly on $N$). The agreement between the numerics and \eref{eq:kscale} is excellent.

\begin{figure}
\begin{center}\includegraphics{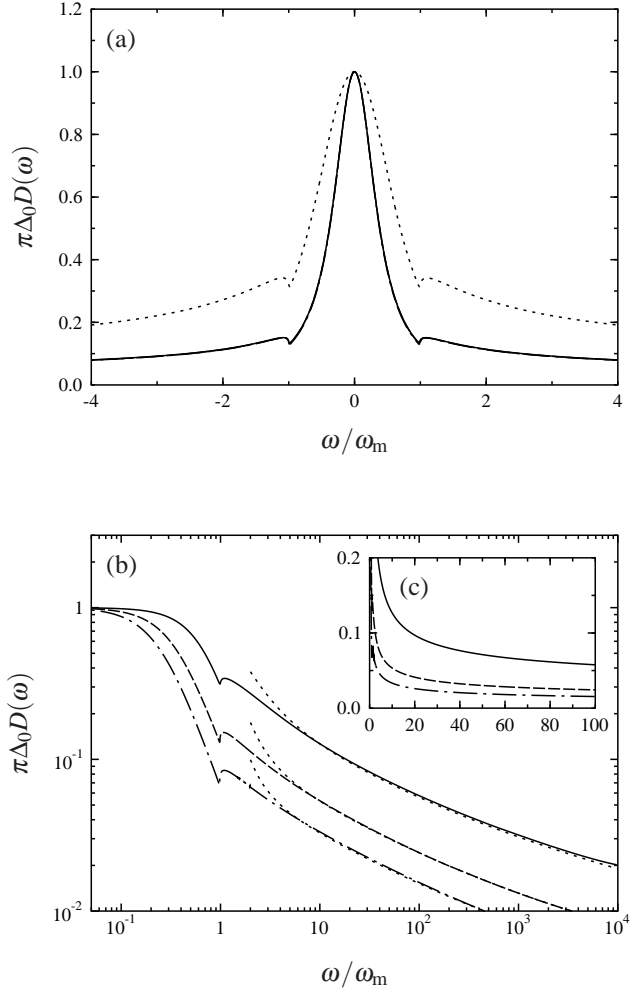}\end{center}
\caption{\label{fig:scalspec} Universal scaling spectra. (a) Scaling collapse of the $SU(4)$ spectrum in strong coupling ($\tilde U \gg 1$), as a function of $\omega/\omegam$, for $\tilde U = 30$, $40$ and $50$ (solid, dashed,
and dot-dashed lines respectively). Spectral collapse is essentially perfect on the energy scales shown. The dotted line shows the corresponding universal spectrum for the $SU(2)$ model \cite{let}. \\ (b) 
Scaling spectra for $SU(2)$, $SU(4)$ and $SU(6)$ (solid, dashed and dot-dashed lines)
on a logarithmic scale, together with comparison (dotted lines) to the analytic forms in \eref{eq:scalspec};  \\ (c) shows the decay of the spectral tails for $SU(2)$, $SU(4)$ and $SU(6)$ on a linear scale.}
\end{figure}

\subsection{Spectral scaling}
\label{sec:specscale}
Having identified the Kondo scale $\omega_\mathrm{m}$, we return to the issue of spectral scaling in more detail.  Figure~\ref{fig:scalspec} shows explicitly the scaling collapse of the $SU(4)$ Kondo resonance as a function of $\omega/\omega_\mathrm{m}$, for $\tilde U=30$, $40$ and $50$ (all collapse perfectly to the scaling form
over the $\omega/\omega_\mathrm{m}$ range shown). For comparison, the LMA $SU(2)$ scaling spectrum \cite{let,nigelscalspec} is also shown, dotted line.
The figure highlights the clear $N$-dependence of the $SU(2N)$ scaling spectrum, alluded to in 
\sref{sec:specoverview}. The $SU(4)$ Kondo scaling resonance is seen to be somewhat narrower in form compared
to its $SU(2)$ counterpart (although it is of course broader on an `absolute' $\omega$-scale, reflecting as discussed
above that  $\omega_\mathrm{m}(N=4) \gg \omega_\mathrm{m}(N=2)$ for given $\util$).

The scaling behaviour of the tails themselves is considered in \fref{fig:scalspec}(b). Here we show the universal forms of $\pi\Delta_0 D(\omega)$ \vs\ $\omega/\omegam$ for the $SU(2)$, $SU(4)$ and $SU(6)$ cases,
plotted with logarithmic axes to highlight the $N$-dependence of the spectral tails (the corresponding
linear plot is shown in \fref{fig:scalspec}(c) for comparison). As $N$ increases, the tails are seen to decay 
more rapidly, although the form of the decay is evidently similar in each case.

This behaviour of the scaling spectrum, for $|\omega|/\omega_\mathrm{m} \gg 1$ and arbitrary $N$, can in
fact be obtained in closed form, following closely the approach of \cite{nigelscalspec}. Details of the calculation are given in \ref{app:scalspec}: the final result is found to be
\begin{equation}
\label{eq:scalspec}
\pi \Delta_0 D(\omega)\overset{\bar\omega\gg 1}\sim\\
\frac{1}{2}\left[ \frac{1}{\left(\frac{4N}{\pi}\ln|\bar\omega|\right)^2 + 1} + \frac{4N+1}{\left(\frac{4N}{\pi}\ln|\bar\omega|\right)^2 + (4N+1)^2}\right]
\end{equation}
where $\bar\omega = \omega/\omega_\mathrm{m}$. The asymptotic results for $N=1$, $2$ and $3$ are shown in \fref{fig:scalspec} as dotted lines, and are seen to agree excellently with the numerical curves for $|\bar\omega| \gtrsim 5$.

\subsection{Comparison with NRG}
We conclude our discussion of the $SU(2N)$ AIM by comparing the scaling spectra obtained above with essentially 
exact numerics from NRG \cite{wilson,kww1,bkp,Weichselbaum,PetersPruschkeAnders}, employing the full density matrix formulation of the method~\cite{PetersPruschkeAnders,Weichselbaum}, and with the self-energy determined directly~\cite{bullahewprus}. At the time of writing, available computational power renders it impossible to obtain accurate NRG results for models with $N>2$. Nevertheless, the LMA scaling spectrum for $N=1$ is known already to agree very well with the NRG result \cite{nigelscalspec}, and here we show that the same is true for $N=2$. As such, it is difficult to imagine the situation changing significantly for higher values of $N$.

\begin{figure}
\begin{center}\includegraphics{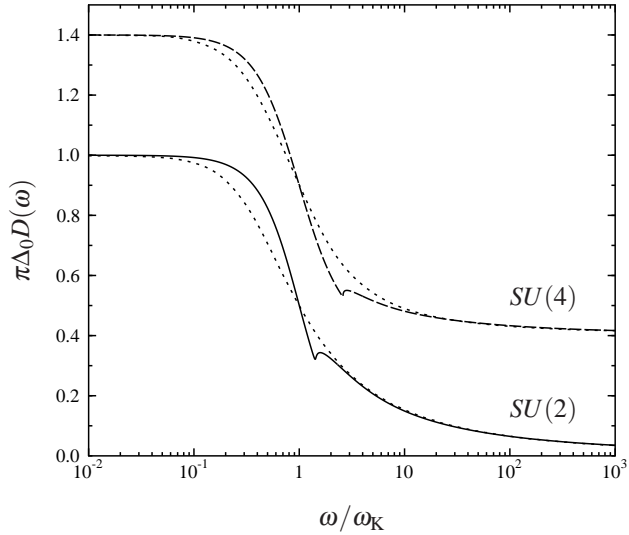}\end{center}
\caption{\label{fig:compnrg} Comparison between LMA and NRG scaling spectra. Solid and dashed lines show respectively the LMA $SU(2)$ and $SU(4)$ scaling spectra, while the dotted lines are corresponding NRG results. 
The Kondo scale $\omega_\mathrm{K}$ is defined by $\pi\Delta_{0}D(\omega_\mathrm{K}) =1/2$. 
Note that the $SU(4)$ spectra have been shifted upward by $0.4$ for clarity.}
\end{figure}

Figure~\ref{fig:compnrg} compares the LMA scaling spectra for the $SU(2)$ and $SU(4)$ models (solid and dashed lines respectively) with those of NRG (dotted); shown as a function of $\omega/\omega_{\mathrm{K}}$ with
$\omega_{\mathrm{K}}$ defined by $\pi\Delta_{0}D(\omega_{\mathrm{K}})=1/2$. Note that
both $SU(4)$ spectra have been shifted upwards by $0.4$ for clarity.

The agreement between the LMA and NRG scaling spectra is striking. The tails of the spectra for $|\omega|/\omega_\mathrm{K}\gg 1$ are essentially indistinguishable, suggesting that the asymptotic form in \eref{eq:scalspec} may be exact. While the LMA and NRG spectra begin to deviate when $\omega/\omega_\mathrm{K}$ is reduced (in part due to the RPA `glitch' in the LMA spectra), symmetry restoration ensures the correct spectral pinning required by the Friedel sum rule, \viz\ $\pi\Delta_0 D(\omega =0) = 1$, and hence the LMA and NRG spectra once again coincide as $\omega\to 0$.

%
%---------------------------------------------------------------------
\section{Away from $SU(2N)$ symmetry}
\label{sec:ccresults}

The discussion of sections~\ref{sec:model}--\ref{sec:results} has naturally focussed on the fully-symmetric $SU(2N)$ Hamiltonian of \eref{eq:su2nhamndots}. We now comment briefly on the effect of an anisotropic interaction between particles on the impurity. Taking \eref{eq:su2nhamndots} and replacing the coefficient of the final term by $U'/2$, \viz 
\begin{equation}
\label{eq:su2nhamndotsasym}
\fl\hat H = \sum_i\left[\sum_{\bk,\sigma}\epsilon_\bk\cre{\bk i\sigma}\des{\bk i\sigma} + \epsilon \hat n_i + U\hat n_{i\uparrow}\hat n_{i\downarrow} + 
V\sum_{\bk,\sigma}\left(\cre{\bk i \sigma}\fdes{i\sigma} + \mathrm{h.c.}\right)\right] +  \frac{U'}{2}\sideset{}{'}\sum_{i,j} \hat n_i \hat n_j,
\end{equation} 
leads to a model of capacitively-coupled $SU(2)$ AIMs with distinct onsite and intersite interactions, $U$ and $U'$ respectively~\cite{ddprl,mrgccstat,mrgccdyn}. For $U'\ne U$, the Hamiltonian has $SU(2)^N$ symmetry rather than the full $SU(2N)$ symmetry of the parent model. We shall focus here on the case $U' < U$ for simplicity.

The consequence of explicit symmetry breaking in the Hamiltonian \eref{eq:su2nhamndotsasym} is that for
$U' < U$, the lowest energy broken symmetry mean-field solutions are now only those with uniform charge on all sites $i$ and broken spin-symmetry. On repeating the arguments of \sref{sec:uhf} however, the structure of the MF Green function for $U' < U$  remains precisely that given in \erefs{eq:mfg1}{eq:mfg2}.
The LMA can then be set up in the same way as described in \sref{sec:lmass}, taking care now to distinguish between $U$ and $U'$ vertices in \fref{fig:sesum}. Since each term in \fref{fig:sesum}(b) involves repeated interactions between $(i,\sigma)$ and the \emph{same} $(j,\sigma')$, each term involves either $U$ \emph{or} $U'$ vertices, not a mixture of the two. It is then readily shown that the LMA self-energy of \eref{eq:sigsum} generalises to
\begin{equation}
\label{eq:sigsumasym}
\Sigma_{i\sigma}(\omega) = \Sigma^{\sigma-\sigma}_{i\sigma}(\omega; U) + (N-1) \left[\Sigma^{\sigma-\sigma}_{i\sigma}(\omega; U') + \Sigma^{\sigma\sigma}_{i\sigma}(\omega; U')\right]
\end{equation} 
where the $U$ or $U'$ vertices appearing in the  various terms are labelled explicitly.
For the reasons discussed in \ref{app:kscale}, the contribution of the $\Sigma^{\sigma\sigma}_{i\sigma}(\omega; U')$ term in \eref{eq:sigsumasym} is negligible when $\tilde U \gg 1$. The two remaining terms by contrast both involve the transverse spin-polarization propagator $\Pi^{\sigma-\sigma}(\omega)$, the resonance in which generates the low-energy Kondo scale within the LMA [see \sref{sec:kscale}]. Neither term can therefore be neglected \emph{a priori}. And since each is calculated with different interaction vertices, there are now \emph{two} distinct energy scales in the problem when $U'\ne U$.

\begin{figure}
\begin{center}\includegraphics{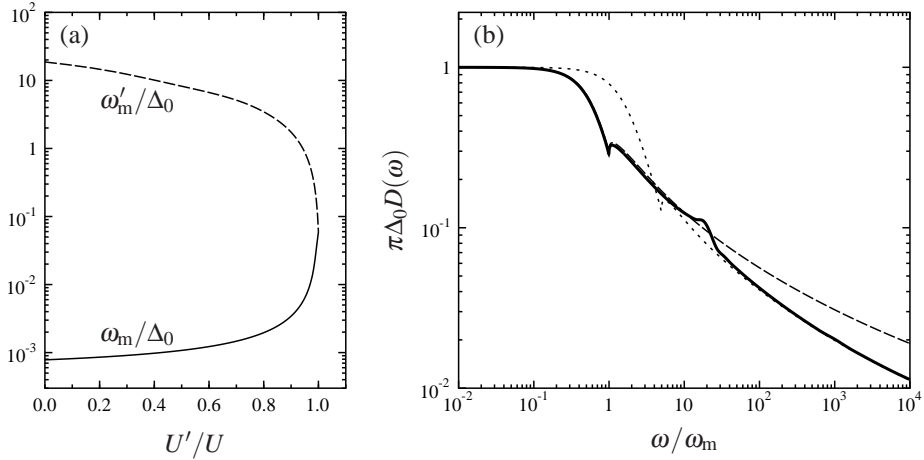}\end{center}
\caption{\label{fig:su4su2spec} Results away from $SU(2N)$ symmetry. (a) The behaviour of the two energy scales $\omegam/\Delta_0$ (solid line) and $\omegam'/\Delta_0$ (dashed line), as a function of $U'/U$, for fixed $\tilde U = 20$ and $N=2$. (b) The universal single particle spectrum $\pi\Delta_0 D(\omega)$ calculated close to the $SU(4)$-symmetric point, with $\alpha = (U-U')/\omega_\mathrm{m}(U'=U) = 5$ (solid line). Also shown is the
pure $SU(2)$ scaling spectrum (dashed line), and the pure $SU(4)$ scaling spectrum (dotted).}
\end{figure}

We have studied the ensuing $U'<U$ behaviour in detail, both numerically and where possible analytically. The typical behaviour of the two energy scales is shown for the illustrative case of $SU(4)$ ($N=2$) with $\tilde U = 20$ in \fref{fig:su4su2spec}(a). We denote by $\omegam$ the scale that enters the first term in \eref{eq:sigsumasym} (arising from spin-flip excitations on the same site), shown in \fref{fig:su4su2spec}(a), while the other scale (due to spin-flip excitations involving distinct sites) is labelled $\omegam'$ and also shown in the figure.

The two-scale behaviour is in good agreement with NRG results \cite{ddprl,mrgccstat}. The low-energy scale $\omegam$ deviates very little from its $U=0$ $SU(2)$ behaviour [$\propto \exp(-\pi \tilde U/8)$, \eref{eq:kscale} with $N=1$]
on initially increasing $U'/U$. Only when $U-U'$ itself becomes of the order of the $SU(4)$ Kondo scale does the scale rapidly cross over to the $U'=U$ behaviour [$\omegam \propto \exp(-\pi \tilde U/16)$] discussed in \sref{sec:kscale} \cite{ddprl}. The higher energy scale $\omegam'$ is typically of order $U-U'$, and arises  in the NRG as a crossover scale from the $SU(4)$ local moment (LM) fixed point to the $SU(2)$ LM fixed point \cite{mrgccstat}. And as $U'\to U$, \fref{fig:su4su2spec}(a) shows that $\omegam'\to \omegam$, reflecting the natural existence of only a single scale at the $SU(4)$ point $U'=U$. 

From our current perspective the most interesting physics is seen when the two scales are comparable; \ie\ when $U-U'\sim {\cal{O}}(\omegam)$. To express this more precisely, we define the quantity
\begin{equation}
\alpha = \frac{U-U'}{\omegam(U'=U)}
\end{equation}
with $\omegam(U'=U)$ the $SU(4)$ Kondo scale arising when $U' =U$. For fixed $N$ we find that when $\tilde U \gg 1$, the ratios $\omegam'/\omegam$  and $\omegam/\omegam(U'=U)=\gamma(\alpha)$ are \emph{universal} functions of $\alpha$ for any finite $\alpha$. As a result, spectra for fixed $\alpha>0$ show universal scaling collapse in the same way as seen previously for $\alpha=0$ (\sref{sec:specscale}).

We show a representative example of one of these scaling spectra in \fref{fig:su4su2spec}(b), for the case $\alpha=5$ (solid line). Note that in addition to the spectral `glitch' at $\omega = \omegam$, there is now another at $\omega=\omegam'\simeq 25\omegam$. Both are a simple consequence of the RPA self-energy approximation (and can be removed if so desired~\cite{nigelscalspec}), although in this case they serve a useful purpose in marking out the low-energy scales of the model. \Fref{fig:su4su2spec}(b) shows in addition the universal $SU(2)$ scaling spectrum as a function of $\omega/\omegam$ (dashed line), and the universal $SU(4)$ spectrum also as a function of $\omega/\omegam = \omega/[\gamma(\alpha)\omegam(U'=U)]$ (dotted line). Just as found in NRG studies~\cite{mrgccdyn},
we see immediately that the $U^{\prime}<U$ scaling spectrum shows a clear crossover from universal $SU(4)$ behaviour when $|\omega|\gg \omegam'$, to universal $SU(2)$ behaviour when $|\omega|\ll\omegam'$, reflecting the different effective low-energy models that arise on the two distinct energy scales~\cite{ddprl,mrgccdyn}.
%
%---------------------------------------------------------------------
\section{Conclusions}
\label{sec:conc}
We have considered here a local moment approach to single-particle dynamics of the orbitally
degenerate $SU(2N)$ Anderson model; focussing explicitly on the particle-hole symmetric case for arbitrary $N$,
where the impurity is occupied by $N$ electrons -- a tangible example being the
middle of the $2$-electron Coulomb blockade valley in a carbon nanotube quantum dot with $SU(4)$ symmetry
\cite{makarovski,choinano,busser,fbaprl}. The resultant LMA is a natural generalisation of the technique developed previously for $SU(2)$ quantum impurity models; and is seen to provide a rather good description of the dynamics of the $SU(2N)$ Anderson model,  recovering the correct exponential vanishing of the Kondo scale for $\tilde U \gg 1$ and its associated universality, the broadening of the Hubbard satellites by many-body scattering processes, and a very good description of the single-particle scaling spectrum  as judged by direct comparison
to NRG results for the $SU(4)$ and $SU(2)$ models. \\ \\

\appendix
\section{Closed form of ${}^0\Pi^{\sigma\sigma}(\omega)$}
\label{app:piform}
The longitudinal spin polarization bubble ${}^0\Pi^{\sigma\sigma}(\omega)$ is calculated from \eref{eq:scrgdef} and \eref{eq:pinoughtdef}. We find 
\begin{eqnarray}
\pi\Delta_0\Re{}^0\Pi^{\sigma\sigma}(\tilde\omega) &= \frac{4 \sigma f(\tilde\omega) - \tilde\omega g(\tilde\omega)}{2h(\tilde\omega)}\label{eq:repiss}\\
\sgn(\omega)\pi\Delta_0\Im{}^0\Pi^{\sigma\sigma}(\tilde\omega) &=\frac{\sigma \tilde\omega f(\tilde\omega) + g(\tilde\omega)}{h(\tilde\omega)}\label{eq:impiss}
\end{eqnarray}
where
\begin{eqnarray}
f(\tilde\omega) &=\tan^{-1}(\tilde\omega+\tilde x) - \tan^{-1}(\tilde\omega-\tilde x)\label{eq:fw}\\
g(\tilde\omega) &= \ln\left\{\frac{\left[(\tilde\omega+\tilde x)^2+1\right]\left[(\tilde\omega-\tilde x)^2+1\right]}{(\tilde x^2+1)^2}\right\}\label{eq:gw}\\
h(\tilde\omega) &= \tilde\omega(\tilde\omega^2+4).\label{eq:hw}
\end{eqnarray}

\section{Derivation of \eref{eq:kscale}}
\label{app:kscale}
The derivation of \eref{eq:kscale} for arbitrary $N$ is closely analogous to that of the $N=1$ case \cite{let}. We begin by noting from \eref{eq:muscssb} that the MF moment $|\mu_0|\to 1^-$ when $\tilde U\to \infty$. Since stability of the RPA polarization propagators requires \eref{eq:stability}, symmetry restoration takes place at an $x\simeq U/2$ and thus, from \eref{eq:mubarx}, $|\bar\mu|\to 1^-$ also. Hence the symmetry restoration condition, \eref{eq:srsym2}, becomes
\begin{equation}
\label{eq:dsfldsfjksdlj}
\Sigma_{i\uparrow}(\omega=0) \overset{\tilde U\gg 1}\sim \frac{U}{2},
\end{equation}
or, equivalently,
\begin{equation}
\label{eq:flddsf}
N\Sigma^{+-}_{i\uparrow}(\omega=0) + (N-1)\Sigma^{++}_{i\uparrow}(\omega=0) \overset{\tilde U\gg 1} \sim\frac{U}{2}
\end{equation}
from \eref{eq:sigsum}.

The first term on the left-hand side of \eref{eq:flddsf} is simply $N$ times the LMA self-energy for the $SU(2)$ AIM, the asymptotic $\tilde U \gg 1$ behaviour of which has already been examined in detail \cite{let}. The key is to note that the RPA $\Pi^{+-}(\omega)$ is dominated by a sharp resonance at $\omega=\omegam$, which 
as $\util \rightarrow \infty$ tends asymptotically to a delta-function, leading to \cite{let}
\begin{equation}
\label{eq:asdf}
\Sigma^{+-}_{i\uparrow}(\omega=0) \overset{\tilde U\gg 1}\sim \frac{4\Delta_0}{\pi}\ln\left(\frac{U}{\omega_\mathrm{m}}\right).
\end{equation} 

The second term on the left-hand side of \eref{eq:flddsf} can be analysed in a similar fashion, starting from the closed form of ${}^0\Pi^{++}(\omega)$ in \ref{app:piform}, and using \eref{eq:rpapi} to obtain the RPA longitudinal spin polarization propagator $\Pi^{++}(\omega)$. Crucially, in marked contrast to the $\Pi^{+-}(\omega)$ discussed above, we find that no sharp resonance develops in the RPA $\Pi^{++}(\omega)$ when $\tilde U\gg 1$. The numerically-determined $\Sigma^{++}_{i\uparrow}(\omega=0)$ thus turns out to be $\mathcal{O}(U^{-1})$, and hence in the limit $\tilde U\gg 1$, $\Sigma^{++}_{i\uparrow}(\omega=0)$ can be neglected in comparison to the logarithmically-diverging $\Sigma^{+-}_{i\uparrow}(\omega=0)$ of \eref{eq:asdf}. Combining \eref{eq:flddsf} and \eref{eq:asdf} then leads to \eref{eq:kscale} straightforwardly.

\section{Derivation of \eref{eq:scalspec}}
\label{app:scalspec}
The tails of the $SU(2N)$ scaling spectra can be obtained in closed form within the LMA. We define $\bar\omega = \omega/\omegam$ and consider the behaviour of the LMA self-energies for fixed $\bar\omega$ in the formal scaling limit $\omegam\to 0$. Let us rewrite \eref{eq:sigsum} explicitly as 
\begin{equation}
\label{eq:sigsumscal}
\Sigma_{i\sigma}(\bar\omega\omegam) = N \Sigma^{\sigma-\sigma}_{i\sigma}(\bar\omega\omegam) + (N-1) \Sigma^{\sigma\sigma}_{i\sigma}(\bar\omega\omegam).
\end{equation}
As described in \ref{app:kscale} the imaginary part of the longitudinal spin polarization propagator, $\Im\Pi^{\sigma\sigma}(\omega)$, contains only broad resonances at $\omega\sim \mathcal{O}(U)$ rather than the emerging delta-function at $\omega=\omegam$ in $\Im\Pi^{\sigma-\sigma}(\omega)$. This means that $ \Sigma^{\sigma\sigma}_{i\sigma}(\omega)$ is \emph{non}-universal at low frequencies and hence, in the limit $\omegam\to 0$ for fixed $\bar\omega$, we can write $ \Sigma^{\sigma\sigma}_{i\sigma}(\bar\omega\omegam)=\Sigma^{\sigma\sigma}_{i\sigma}(0)$. But since $\Sigma^{\sigma\sigma}_{i\sigma}(0) = \mathcal{O}(U^{-1})$ (as in \ref{app:kscale}), it vanishes in the formal scaling limit $\tilde U\to\infty$. Hence \eref{eq:sigsumscal} reduces simply to
\begin{equation}
\label{eq:sigsumscal2}
\Sigma_{i\sigma}(\bar\omega\omegam) \overset{\omegam\to0}= N \Sigma^{\sigma-\sigma}_{i\sigma}(\bar\omega\omegam).
\end{equation}

Moreover, $\Sigma^{\sigma-\sigma}_{i\sigma}(\bar\omega\omegam)$ is precisely the function already analysed to obtain the scaling spectrum of the $SU(2)$ AIM \cite{nigelscalspec}. All the analysis of \cite{nigelscalspec} can thus be generalised to the $SU(2N)$ model quite straightforwardly. The upshot is that eqn.~(5.1) of \cite{nigelscalspec} must simply be multiplied by $N$ (reflecting the additional factor of $N$ in \eref{eq:sigsumscal2}) which, by analogy to eqn.~(5.2) of \cite{nigelscalspec}, gives the full LMA scaling spectrum for the $SU(2N)$ AIM as
\begin{equation}
\label{eq:scalspecfull}
\fl\pi \Delta_0 D(\omega)=
\frac{1}{2}\Biggl[ \frac{1}{\left[\frac{4N}{\pi}\ln|\bar\omega + 1|\right]^2 + 1} + \frac{1+4N\theta(\bar\omega-1)}{\left[\frac{4N}{\pi}\ln|\bar\omega-1|\right]^2 + \left[1+4N\theta(\bar\omega-1)\right]^2}\Biggr]
\end{equation}
for $\bar\omega > 0$ (the corresponding negative-frequency spectrum follows from particle-hole symmetry, \ie\ $D(-\omega)=D(\omega)$). In the limit $|\bar\omega|\gg1$,  \eref{eq:scalspec} results.

\ack
We are grateful for financial support from the EPSRC, under grant EP/D050952/1. MRG acknowledges the use of the UK National Grid Service, on which some of the NRG calculations for this work were performed.

%---------------------------------------------------------------------
\bibliographystyle{iopart-num}
\section*{References}
\bibliography{paper}
\end{document}